\documentclass[10pt,letterpaper,twocolumn]{article} %% two column, final layout

\usepackage{ol2}
\usepackage[draft]{hyperref}
\usepackage{amsmath}

\begin{document}

\twocolumn[ % activate for two-column option

\title{In-vivo two-photon imaging of the honey bee antennal lobe}

%% For REVTeX it is possible to automate superscript and e-mail callouts with the superscriptaddress option; see REVTeX4 documentation.

\author{Albrecht Haase,$^{1,*}$ Elisa Rigosi,$^{2,3}$ Federica Trona,$^2$ Gianfranco Anfora,$^2$ Giorgio Vallortigara,$^3$\\ Renzo Antolini,$^1$ and Claudio Vinegoni$^{1,4}$}

\address{
$^1$Physics Department, University of Trento, \\ via Sommarive 14, 38050 Povo (TN), Italy\\
$^2$IASMA Research and Innovation Center, Fondazione E. Mach, \\ via E. Mach 1, 38010 San Michele all'Adige, Italy\\
$^3$Center for Mind/Brain Sciences, University of Trento, \\ Corso Bettini 31, 38068 Rovereto, Italy\\
$^4$Center for Systems Biology, Massachusetts General Hospital, Harvard Medical School, \\ 185 Cambridge Street, Boston, Massachusetts 02114, USA\\
$^*$Corresponding author: albrecht.haase@unitn.it
}

\begin{abstract}
Due to the honey bee's importance as a simple neural model, there is a great need for new functional imaging
modalities. Herein we report on the use of two-photon microscopy for in-vivo functional and morphological imaging
of the honey bee's olfactory system focusing on its primary centers, the antennal lobes (ALs). Our imaging platform
allows for simultaneously obtaining both morphological measurements of the AL and in-vivo calcium recording of
neural activities. By applying external odor stimuli to the bee's antennas, we were able to record the
characteristic odor response maps. Compared to previous works where conventional fluorescence microscopy
is used, our approach  offers all the typical advantages of multi-photon imaging, providing substantial
enhancement in both spatial and temporal resolutions while minimizing photo-damages and autofluorescence
contribution with a four-fold improvement in the functional signal. Moreover, the multi-photon associated extended penetration depth allows for functional imaging within profound glomeruli.
\end{abstract}

] %% activate for two-column option

\bigskip
%Introduction
Important advances in neuroscience have always been strongly linked to the development of new investigative tools.
While activity in single neurons has been detected for the first time by electrical recording
\cite{Hodgkin1939}, later on, the development of a new class of voltage-sensitive dyes \cite{Salzburg1973} has
offered the possibility to optically image the functionality of neuronal circuits at both the single neuron and
whole brain levels. In recent years the development of calcium-sensitive dyes \cite{Grynkiewicz1985} has provided
another universal and sensitive method to study distinct information processing pathways in whole neural networks
in particular when used in combination with two-photon laser scanning microscopy\cite{Denk1990}, allowing for
in-vivo real-time monitoring of complex neural circuits down to several hundred micrometers within the specimen
\cite{Svoboda1997}. In this paper we focus our attention on two-photon calcium imaging of the honey bee's ({\it
Apis mellifera}) brain. With less than one million neurons, the honey bee is an excellent model for the study of
neural systems of intermediate complexity \cite{Menzel2001} making it an ideal candidate for two-photon microscopy
\cite{Franke2009}.
\\
In the past years several optical imaging modalities have allowed to gain tremendous insights into the bee's olfactory system. After early pioneering works using voltage-sensitive dyes \cite{Lieke1993}, various staining techniques have enabled to investigate different aspects of the odor processing network by fluorescence microscopy. In-vivo experiments using calcium-sensitive cell-permeant dyes have  visualized the activity patterns of the glomeruli, the AL's functional units\cite{Galizia1999}. Signals were dominated by the Olfactory Receptor Neurons (ORNs) which are the input channels to the AL. Selective backfill staining with membrane-impermeable dyes has instead allowed to record the AL's output signal from the Projection Neurons (PNs) \cite{Sachse2002}. In addition to the functional properties, also the morphology of the olfactory centers has been studied extensively, mostly by confocal microscopy \cite{Brandt2005}, and evidence of glomerular plasticity in the ALs has been recently discovered\cite{Hourcade2009}. While linear macro- and microscopy imaging techniques have been proven to be extremely successful in order to characterize this complex neuronal system, their intrinsic limitations have become more and more obvious \cite{Galizia2001}. Full-field microscopy does not offer sufficient axial resolution to resolve the exact origin of functional signals in 3D and lacks the temporal resolution to determine whether valuable information might be encoded in the temporal structure of the recorded odor-evoked signals. Whereas confocal microscopy due to its intrinsic photo-damaging properties poses severe time constraints to in-vivo imaging sessions and has therefore only be used for morphological studies of the extracted brain.

Herein we report on neural imaging of the honey bee's antennal lobe by two-photon microscopy to overcome the imaging impediments currently faced by neuroscientists. Our imaging platform enables to acquire both in-vivo functional and morphological data of the ALs. Functional data show that we have been able to spatially resolve firing neurons deep within the AL, while the high temporal resolution permits a reconstruction of the neuron's firing rate \cite{Moreaux2007}. The intrinsic two-photon optical penetration is deep enough to offer the possibility to study, beside the well investigated T1 glomeruli which are projecting into the lateral antenno-cerebralis tract (l-ACT), in the future classes of neurons that have not been accessible in the ALs yet, namely the T2, T3, and T4 glomeruli projecting into the medial antenno-cerebralis tract (m-ACT) \cite{Yamagata2009}.

%Methods
Bees have been prepared in accordance to a well established protocol \cite{Galizia2004}. After being fixed to a custom made imaging stage, a small window was cut into the head's cuticula above the mushroom body, glands and trachea were gently moved aside, and a solution of calcium sensitive dye (fura2-dextran, Invitrogen with BSA, Sigma-Aldrich) was injected into the antenno-cerebralis tracts below the $\alpha$-lobe. Then the cuticula was carefully closed and the animals were stored for 20h in order for the dye to diffuse into the AL. Before the imaging session, the cuticula, the glands, and the trachea above the AL were removed.

\begin{figure}[htb] 
\centerline{\includegraphics[width=8cm]{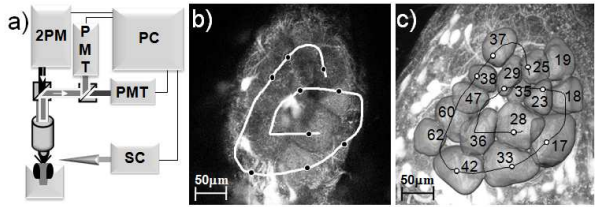}} 
\caption{a) Experimental setup: The exposed AL of a bee is imaged with an upright two-photon microscope (2PM). The epifluorescence signal is collected by way of a dichroic mirror and detected by a photomultiplier tube (PMT). A stimulus controller (SC)produces time-gated odor puffs and triggers signal recording. (b) Image of a right antennal lobe at 25$\mu$m depth: The line indicates the laser scanning trace, the dots label the measurement's reference positions corresponding to the vertical lines in Figure \ref{fig:2}. (c) Projection view of the AL volume image stack, superimposed by the reconstructed surface plots of the involved T1 glomeruli, identified and labeled according to \cite{Menzelweb}.}
\label{fig:1}
\end{figure}

The experimental setup, sketched in Figure~\ref{fig:1}a, consists of an upright two-photon microscope (Ultima IV, Prairie Technologies) combined with an ultra-short pulsed laser (Mai Tai, Deep See HP, Spectra-Physics) tuned to 800nm for fura-2 excitation. The fluorescence signal was detected after a 70nm band-pass filter centered at around 525nm. All images were acquired with a water immersion objective (40x, NA 0.8, Olympus). Optimal signal-to-noise ratio was achieved with laser powers of about 10mW on the sample without observing any induced photo-damage. The maximum penetration depth for morphological imaging was found to be 400$\mu$m, while functional signals could be recorded down to a depth of 150$\mu$m within the ALs. A high functional temporal resolution of about 15ms was obtained by laser scanning along one-dimensional custom-defined traces, crossing the glomeruli of interest through an arbitrary horizontal plane (Fig.~\ref{fig:1}b). All acquired data have been corrected for photo-bleaching, while 2D running-average filtering was used to reduce the noise level. Spatial averaging was performed over a typical glomerulus size of 30$\mu$m (compare Fig.~\ref{fig:1}c), while temporal averaging was applied over 80ms preserving all main dynamic features of the data. A stimulus controller (CS-55, Syntech) delivered odor stimuli to the antennas in pulses of 2s length without changing the total air flux and triggered the microscope recording 3s before odor delivery. The total optical recording cycle length was 9s.

\begin{figure}[htb]
\centerline{\includegraphics[width=8cm]{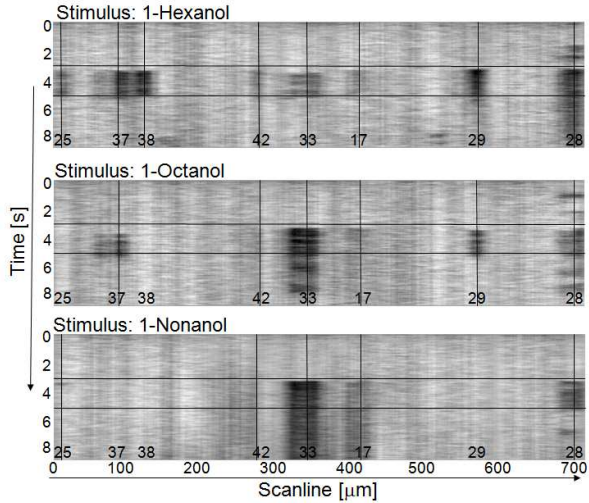}} \caption{Calcium response maps for three different odors (1-Hexanol above, 1-Octanol middle, 1-Nonanol below), recorded along the scanning trace in Figure\ref{fig:1} (b,c).  The stimulus period is enclosed by the horizontal lines, responding glomeruli centers are marked by vertical lines, numbers label the identified glomeruli.} \label{fig:2}
\end{figure}

We first recorded the spatio-temporal functional activity in the AL by measuring the two-photon calcium response
signal along the line traces indicated in Figure~\ref{fig:1}(b,c) for stimuli triggered by three different floral components:
1-Hexanol, 1-Octanol, and 1-Nonanol. Enhanced neural activity, leading to an increasing intra-neuronal calcium concentration, causes a drop in the measured two-photon fluorescence intensity, producing dark bands in the scanlines-over-time maps at the positions of the corresponding glomeruli. We detected response signals of up to 20\% intensity change, which is about 4 times higher than in comparable experiments using wide field imaging. The recorded maps are shown in Figure~\ref{fig:2} and reproduce features which have already been observed by conventional single-photon fluorescence microscopy \cite{Galizia1999}, such as the very broad response of glomeruli T1-17, T1-28, and T1-33 to all tested odors. Likewise, 1-Hexanol has been found to produce responses in several of the monitored glomeruli. Strikingly different from previously published data obtained with full-field microscopy \cite{Peele2006} are the quite strong responses of glomeruli T1-29 and T1-37 for both 1-Hexanol and 1-Octanol.

\begin{figure}[htb] \centerline{\includegraphics[width=8cm]{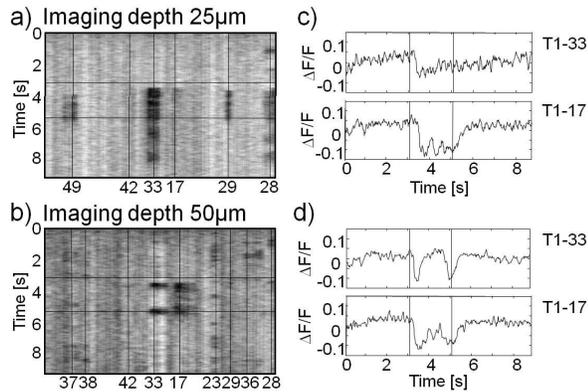}} \caption{Odor response maps at depths of 25$\mu$m (a) and 50$\mu$m (b) within the AL. The signal is plotted as a function of position along the line trace indicated in Figure~\ref{fig:1} (x-axis) and as a function of time (y-axis). The stimulus period is enclosed by horizontal lines, the single responding glomeruli are marked by vertical lines. (c,d) Single temporal traces for the strongest responding glomeruli T1-33 and T1-17 at the two corresponding depths.} \label{fig:3}
\end{figure}

We have then exploited the larger penetration depth and the higher axial resolution offered by our setup, in order to obtain functional spatio-temporal odor response maps at different axial positions within
the AL. Figure~\ref{fig:3} shows the calcium response maps to an 1-Octanol odor stimulus at a depth of 25$\mu$m (Fig.~\ref{fig:3}a) and 50$\mu$m (Fig.~\ref{fig:3}b) respectively. The high axial resolution allows to clearly resolve the functional activity at the different depths. In particular the responses of the upper surface glomeruli T1-37 and T1-29, clearly visible at the imaging depth of 25$\mu$m, disappear at 50$\mu$m,  while the weak response of the glomerulus T1-17 at 25$\mu$m becomes more pronounced at 50$\mu$m. The temporal traces of the single glomeruli data (Fig.~\ref{fig:3}c,d) allow to analyze possible temporal components of the olfactory code, like response delay or oscillatory responses as reported in other animals\cite{Laurent1996}.

%Results

%Discussion
The two-photon microscopy functional data we have presented in this work suggest that this imaging modality
offers the capability to extend the specific AL odor response maps that have been measured in the past for many
different odor components and in most of the T1 glomeruli \cite{Galizia1999}. So far these maps
contained only the static parameters response strength and consistency range, which might now be supplemented by
adding temporal features. The intrinsic axial resolution and the extended imaging depth of two-photon microscopy
has allowed us to resolve profound functional data. Odor response maps could therefore be
completed in the future by measuring the more profound glomeruli of the m-ACT tract.
This possibility is of special interest because glomeruli projecting into the m-ACT have been hypothesized to show fundamentally different properties, e.g. regarding memory related plasticity after odor conditioning \cite{Peele2006}. In addition we have obtained a 4-fold increase in the functional-related fluorescence change with respect to similar experiments using wide field imaging. Another promising feature of a two-photon microscopy approach is the
possibility to investigate sub-glomerular structures down to single neurons \cite{Franke2009}. This becomes even
more crucial if imaging is extended to higher order brain structures such as the mushroom body, where a
meta-structure comparable to the AL's glomeruli is absent. In addition to resolution's improvements, the intrinsic
two-photon limited photo-damage offers extended imaging sessions of several hours. This should allow in the future for in-vivo
morphology studies to monitor in real-time volume plasticity after odor conditioning \cite{Hourcade2009}.

\bigskip
We wish to thank T. Franke and R. Menzel for helpful discussions. C.V. acknowledges funding from the Provincia
autonoma di Trento (project COMNFI).


\begin{thebibliography}{99}
\bibitem{Hodgkin1939}
A. L. Hodgkin and A. F. Huxley,
%"Action potentials recorded from inside a nerve fibre,"
Nature \textbf{144}, 710-711 (1939).
\bibitem{Salzburg1973}
B. M. Salzberg, L. B. Cohen, and H. V. Davila,
%"Optical Recording of Impulses in Individual Neurones of an Invertebrate Central Nervous System,"
Nature \textbf{246}, 508-509 (1973).
\bibitem{Grynkiewicz1985}
G. Grynkiewicz, M. Poenie, and R. Y. Tsien,
%"A New Generation of Ca2+ Indicators with Greatly Improved Fluorescence Properties,"
J. Biol. Chem. \textbf{260}, 3440-3450 (1985).
\bibitem{Denk1990}
W. Denk, J. Strickler, and W. Webb,
%"Two-photon laser scanning fluorescence microscopy,"
Science \textbf{248}, 73-76 (1990).
\bibitem{Svoboda1997}
K. Svoboda, W. Denk, D. Kleinfeld, and D. W. Tank,
%"In vivo dendritic calcium dynamics in neocortical pyramidal neurons,"
Nature \textbf{385}, 161-165 (1997).
\bibitem{Menzel2001}
R. Menzel and M. Giurfa
%"Cognitive architecture of a mini-brain: the honeybee,"
Trends Cognit. Sci. \textbf{5}, 62-71 (2001).
\bibitem{Franke2009}
T. Franke,
%"In vivo 2-photon calcium imaging of olfactory interneurons in the honeybee antennal lobe,"
Dissertation, FB Biologie, Chemie, Pharmazie, Freie Universit\"{a}t Berlin (2009).
\bibitem{Lieke1993}
E. E. Lieke,
%"Optical recording of neuronal activity in the insect central nervous system: odorant coding by the antennal lobes of honeybees,"
Eur. J. Neurosci. \textbf{5}, 49–55 (1993).
\bibitem{Galizia1999}
C.G. Galizia, S. Sachse, A. Rappert, R. Menzel,
%"The glomerular code for odor representationis species specific in the honeybee Apis mellifera,"
Nat. Neurosci. \textbf{2}, 473-478 (1999).
\bibitem{Sachse2002}
S. Sachse and C. G. Galizia,
%"The role of inhibition for temporal and spatial odor representation in olfactory output neurons: a calcium imaging study,"
J. Neurophysiol. \textbf{87}, 1106-1117 (2002).
\bibitem{Brandt2005}
R. Brandt, T. Rohlfing, J. Rybak, S. Krofczik, A. Maye, M. Westerhoff, H.-C. Hege, and R. Menzel,
%"Three-dimensional average-shape atlas of the honeybee brain and its applications,"
J. Comp. Neurol. \textbf{492}, 1-19 (2005).
\bibitem{Hourcade2009}
B. Hourcade, E. Perisse, J.M. Devaud, and J. C. Sandoz,
%"Long-term memory shapes the primary olfactory center of an insect brain,"
Learn. Mem. \textbf{16} 607-615 (2009).
\bibitem{Galizia2001}
C. G. Galizia and R. Menzel,
%"The role of glomeruli in the neural representation of odours: results from optical recording studies,"
J. Insect Physiol. \textbf{47}, 115-130 (2001).
\bibitem{Moreaux2007}
L. Moreaux and G. Laurent,
%"Estimating firing rates from calcium signals in locust projection neurons in vivo,"
Front. Neural Circuits \textbf{1}, 2 (2007).
\bibitem{Yamagata2009}
N. Yamagata, M. Schmuker, P. Szyszka, M. Mizunami, and R. Menzel,
%"Differential odor processing in two olfactory pathways in the honeybee,"
Front. Syst. Neurosci. \textbf{3}, 16 (2009).
\bibitem{Galizia2004}
C. G. Galizia and R. Vetter,
%"Optical methods for analyzing odor-evoked activity in the insect brain,"
in Advances in Insect Sensory Neuroscience, T. A. Christensen, ed. (CRC press, 2004) pp. 349-392.
\bibitem{Menzelweb}
J. Rybak and R. Menzel, "Honeybee antennal lobe atlas", ftp://www.neurobiologie.fu-berlin.de/honeybeeALatlas.
\bibitem{Peele2006}
P. Peele, M. Ditzen, R. Menzel, and C. G. Galizia,
%"Appetitive odor learning does not change olfactory coding in a subpopulation of honeybee antennal lobe neurons,"
J. Comp. Physiol. A Neuroethol. Sens. Neural Behav. Physiol. \textbf{192}, 1083-1103 (2006).
\bibitem{Laurent1996}
G. Laurent, M. Wehr, K. MacLeod, M. Stopfer, B. Leitch, and H. Davidowitz,
%"Dynamic encoding of odors with oscillating neuronal assemblies in the locust brain,"
Biol. Bull. \textbf{191}, 70-75 (1996).
\end{thebibliography}
\end{document}